%%%%%%%%%%%%%%%%%%%%%%%%%%%%%%%%%%%%%%%%%%%%%%%%%%%%%%%%%%%%%%%%%
\input amstex
\magnification =\magstep1
\documentstyle{amsppt}
\pageheight{24truecm}
\pagewidth{16truecm}
\parindent=10pt
\nologo
\TagsOnRight
\topmatter
\font\twtt= cmbx12
\font\ttt=cmbx10
\font\sc=cmcsc10

\title\nofrills
{\twtt Asymptotics of eigenvalues of the Schr{\"o}dinger operator
 with a strong $\delta$-interaction on a loop}
\endtitle
\author
\leftline{{}\hskip .9cm{\bf Pavel Exner}}
\leftline{{}\hskip .9cm{\rm Nuclear Physics
 Institute, ASCR,}}
\leftline{{}\hskip .9cm{\rm 25068 {\v R}e{\v z}
 near Prague, Czech Republic}}
\smallskip
\leftline{{}\hskip .9cm{\rm Doppler Institute, Czech Technical
 University,}}
\leftline{{}\hskip .9cm{\rm B\v{r}ehov\'{a} 7, 11519 Prague,
 Czech Republic}}
\leftline{{}\hskip .9cm{\rm (exner\@ujf.cas.cz)}}
\leftline{}
\leftline{{}\hskip .9cm{\bf Kazushi Yoshitomi}}
\leftline{{}\hskip .9cm{\rm Graduate School of
 Mathematics, Kyushu University,}}
\leftline{{}\hskip .9cm{\rm Hakozaki,
 Fukuoka 812-8581, Japan}}
\leftline{{}\hskip .9cm
{\rm (yositomi\@math.kyushu-u.ac.jp)}}
\endauthor
\abstract\nofrills
{\ttt Abstract}
\smallskip
\noindent In this paper we
 investigate the operator
 $H_{\beta}=-\Delta-\beta\delta(\cdot-\Gamma)$
 in $L^{2}({\Bbb R}^{2})$, where $\beta>0$
 and $\Gamma$ is a closed $C^{4}$
 Jordan curve in ${\Bbb R}^{2}$.
 We obtain the asymptotic form of each eigenvalue
 of $H_{\beta}$ as $\beta$ tends to infinity. We
 also get the asymptotic form of the number of
 negative eigenvalues of $H_{\beta}$ in the strong coupling
 asymptotic regime.
\endabstract
\rightheadtext{Strong delta-interaction}
\leftheadtext{P.Exner and K. Yoshitomi}
\endtopmatter
\redefine\leq{\leqslant}
\redefine\geq{\geqslant}
\document
\leftline{\bf 1. Introduction}
\leftline{}
\noindent In this paper we study the Schr{\"o}dinger operator
 with a $\delta$-interaction on a loop.
 Let $\Gamma:[0,L]\owns s\mapsto
(\Gamma_{1}(s),\Gamma_{2}(s))\in {\Bbb R}^{2}$ be a
 closed $C^{4}$ Jordan curve which is parametrized by
 the arc length. Let $\gamma:[0,L]\rightarrow{\Bbb R}^{2}$
 be the signed curvature of $\Gamma$. For
 $\beta>0$, we define
$$q_{\beta}(f,f)=\Vert\nabla f\Vert^{2}_{L^{2}({\Bbb R}^{2})}
-\beta\int_{\Gamma}|f(x)|^{2}\,dS
\quad\text{for}\quad f\in H^{1}({\Bbb R}^{2}).\tag 1.1$$
By $H_{\beta}$ we denote the self-adjoint operator
 associated with the form $q_{\beta}$. The operator
 $H_{\beta}$ is formally written as $-\Delta-\beta
\delta (\cdot-\Gamma)$.
 Since
 $\Gamma$ is compact in ${\Bbb R}^{2}$, we have
 $\sigma_{\roman{ess}}(H_{\beta})=[0,\infty )$ by [3,
 Theorem 3.1]. Our main purpose is to study the
 asymptotic behaviour of the negative eigenvalues of
 $H_{\beta}$ as $\beta$ tends to infinity. We define
$$S=-\frac{d^{2}}{d s^{2}}-\frac{1}{4}\gamma(s)^{2}
\quad\text{in}\quad
L^{2}((0,L))\tag 1.2$$
with the domain
$$P=\{\varphi\in H^{2}((0,L));\quad
\varphi(L)=\varphi(0),\quad \varphi^{\prime}(L)
=\varphi^{\prime}(0)\}.\tag 1.3$$
For $j\in {\Bbb N}$, we denote by $\mu_{j}$
 the $j$th eigenvalue of $S$ counted with multiplicity. Our
 main results are the following.
\proclaim{\sc Theorem 1} Let $n$ be an arbitrary integer.
 There exists $\beta(n)>0$ such that
$$\sharp\sigma_{\roman d}(H_{\beta})\geq n
\quad\text{for}\quad
\beta\geq \beta(n).$$
For $\beta\geq \beta(n)$
 we denote by $\lambda_{n}(\beta)$ the $n$th
 eigenvalue of $H_{\beta}$ counted
 with multiplicity. Then
 $\lambda_{n}(\beta)$ admits an
 asymptotic expansion of the form
$$\lambda_{n}(\beta)=-\frac{1}{4}\beta^{2}
+\mu_{n}+{\Cal O}(\beta^{-1}\log\beta)
\quad\text{as}\quad\beta
\rightarrow\infty.\tag 1.4$$
\endproclaim
\proclaim{\sc Theorem 2} The function
 $\beta\mapsto\sharp\sigma_{\roman d}(H_{\beta})$ admits
 an asymptotic expansion of the form
$$\sharp\sigma_{\roman d}(H_{\beta})
=\frac{L}{2\pi}\beta+{\Cal
O}(\log\beta)\quad\text{as}
\quad\beta\rightarrow\infty.\tag 1.5$$
\endproclaim
The Schr{\"o}dinger operator with a singular interaction has
 been studied by numerous authors (see [1-3] and the references
 therein). The basic concepts of the theory are summarized in the
 monograph [1]. A particular case of a $\delta$-interaction
 supported by a curve attracted much less attention (see [3, 4]
 and a recent paper [5]). In [3] some upper bounds to the number
 of eigenvalues for a more general class of operators (with $\beta$
 dependent on the arc length parameter) were obtained by the
 Birman-Schwinger argument (see [3, Theorems 3.4,
 3.5, and 4.2]). As it is usually the case with the
 Birman-Schwinger technique, these bounds are sharp for
 small positive $\beta$ (see [3, Example 4.1]) while they give
 a poor estimate in the semiclassical regime. On the contrary,
 our estimate $(1.5)$ is close to optimal for large positive
 $\beta$.
 Our main tools to prove
 Theorem 1 and Theorem 2 are the Dirichlet-Neumann bracketing
 and approximate operators with separated variables.
\newline
\leftline{}
\leftline{\bf 2. Proof of Theorem 1}
\leftline{}
\noindent Let us prepare some quadratic
 forms and operators which we need in the sequel. For this
 purpose,
 we first need the following result.
\proclaim{\sc Lemma 2.1} Let $\Phi_{a}$ be the map
$$[0,L)\times (-a,a)\owns (s,u)\mapsto
(\Gamma_{1}(s)-u\Gamma^{\prime}_{2}(s),
\Gamma_{2}(s)+u\Gamma^{\prime}_{1}(s))
\in {\Bbb R}^{2}.$$
 Then there exists $a_{1}>0$ such that
 the map $\Phi_{a}$ is injective for any $a\in (0,a_{1}]$.
\endproclaim
\demo{Proof} We extend $\Gamma$ to a periodic function
 with period $L$, which we denote by $\tilde{\Gamma}(s)
=(\tilde{\Gamma}_{1}(s),\tilde{\Gamma}_{2}(s))$.
 Since $\Gamma$ is a closed $C^{4}$ Jordan curve, we have
 $\tilde{\Gamma}\in C^{4}({\Bbb R})$.
 We extend $\gamma$ to a function $\tilde{\gamma}$ on ${\Bbb R}$
 by using the formula
 $\tilde{\gamma}(s)=
\tilde{\Gamma}^{\prime\prime}_{1}(s)
\tilde{\Gamma}^{\prime}_{2}(s)
-\tilde{\Gamma}^{\prime\prime}_{2}(s)
\tilde{\Gamma}^{\prime}_{1}(s)$.
 Then $\tilde{\gamma}(\cdot)$
 is periodic with period $L$ and
 $\tilde{\gamma}\in C^{2}({\Bbb R})$.
 By $\Phi$ we denote the map
$${\Bbb R}^{2}\owns (s,u)\mapsto
(\tilde{\Gamma}_{1}(s)-u
\tilde{\Gamma}^{\prime}_{2}(s),
\tilde{\Gamma}_{2}(s)+u
\tilde{\Gamma}^{\prime}_{1}(s))\in {\Bbb R}^{2}.$$
Let $J\Phi$ be the Jacobian matrix of $\Phi$. We put
$$\gamma_{+}=\max_{[0,L]}|\gamma(\cdot)|.$$
We have
$$\det J\Phi(s,u)=1+u
\tilde{\gamma}(s)\geq\frac{1}{2}
\quad\text{for}\quad
(s,u)\in{\Bbb R}\times [-{\tsize\frac{1}{2\gamma_{+}}},\,
{\tsize\frac{1}{2\gamma_{+}}}].\tag 2.1$$
In addition, there exists a constant $M>0$ such that
$$|\partial^{\alpha}_{y}\Phi_{j}(y)|
\leq M\quad\text{on}\quad{\Bbb R}\times
 [-{\tsize\frac{1}{2\gamma_{+}}},\,
{\tsize\frac{1}{2\gamma_{+}}}]\tag 2.2$$
for any $1\leq |\alpha|\leq 2$ and $j=1,2$,
 where $y=(s,u)$
 and $\Phi(y)=(\Phi_{1}(y),\Phi_{2}(y))$.
 Combining [8, Lemma 3.6] with
 $(2.1)$ and $(2.2)$, we claim that
 there exists $a_{0}\in (0,\frac{1}{2\gamma_{+}})$
 such that $\Phi$ is injective on
 $[k-a_{0},k+a_{0}]\times[-a_{0},a_{0}]$
 for all $k\in{\Bbb R}$. We put
$$\tau=\min_{p\in [a_{0},L/2]}\min_{t\in [0,L]}
|\tilde{\Gamma}(t)-\tilde{\Gamma}(t+p)|.\tag 2.3$$
Since $\tilde{\Gamma}$ is injective on $[0,L)$
 and $\tilde{\Gamma}(\cdot)$
 has period $L$, we have $\tau>0$. Put
 $a_{1}=\min\{a_{0},\tau/4\}$. Let us show that
 $\Phi$ is injective on $[0,L)\times(-a_{1},a_{1})$.
 We first prove
 the following claim.

\noindent
(i) {\it Assume that $\Phi(s_{1},u_{1})
=\Phi(s_{2},u_{2})$,
 $|s_{1}-s_{2}|\leq\frac{L}{2}$, and
 $(s_{1},u_{1}),\, (s_{2},u_{2})\in
{\Bbb R}\times(-a_{1},a_{1})$.
 Then we have $(s_{1},u_{1})
=(s_{2},u_{2})$.}

Since $\Phi(s_{1},u_{1})=\Phi(s_{2},u_{2})$
 and $|\tilde{\Gamma}_{j}^{\prime}(\cdot)|\leq 1$
 on ${\Bbb R}$ for
 $j=1,2$, we obtain
$$|\tilde{\Gamma}_{1}(s_{1})-
\tilde{\Gamma}_{1}(s_{2})|
=|u_{1}\tilde{\Gamma}_{2}^{\prime}(s_{1})
-u_{2}\tilde{\Gamma}_{2}^{\prime}(s_{2})|
\leq 2a_{1},$$
$$|\tilde{\Gamma}_{2}(s_{1})
-\tilde{\Gamma}_{2}(s_{2})|
=|u_{1}\tilde{\Gamma}_{1}^{\prime}(s_{1})
-u_{2}\tilde{\Gamma}_{1}^{\prime}(s_{2})|
\leq 2a_{1}.$$
So we have $|\tilde{\Gamma}(s_{1})
-\tilde{\Gamma}(s_{2})|
\leq 2\sqrt{2}a_{1}$, and therefore
$$|\tilde{\Gamma}(s_{1})-\tilde{\Gamma}(s_{2})|<\tau.$$
This together with $(2.3)$ implies that
 $|s_{1}-s_{2}|<a_{0}$. Since $\Phi$ is injective on
 $[s_{1}-a_{0},s_{1}+a_{0}]\times[-a_{0},a_{0}]$
 and $\Phi(s_{1},u_{1})=\Phi(s_{2},u_{2})$, we get
 $(s_{1},u_{1})=(s_{2},u_{2})$.
 In this way we proved $({\roman i})$.

Next we shall prove the following implication.

\noindent
(ii) {\it Assume that $\Phi(s_{1},u_{1})
=\Phi(s_{2},u_{2})$, $s_{1}\leq s_{2}$, and
 $(s_{1},u_{1}),\, (s_{2},u_{2})\in
[0,L)\times(-a_{1},a_{1})$.
 Then we have $s_{2}-s_{1}\leq \frac{L}{2}$.}

We prove this by contradiction. Assume that
 $s_{2}-s_{1}>\frac{L}{2}$. We put $s_{3}=s_{2}-L$. Then we
 get $0<s_{1}-s_{3}<\frac{L}{2}$ and $\Phi(s_{3},u_{2})
=\Phi(s_{1},u_{1})$. As in the proof of $({\roman i})$
 we obtain $(s_{1},u_{1})=(s_{3},u_{2})$
 which violates the fact that $0<s_{1}-s_{3}<\frac{L}{2}$, so
 we proved $(\roman{ii})$.

Combining $({\roman i})$ with $(\roman{ii})$,
 we conclude that $\Phi$ is injective on
 $[0,L)\times(-a_{1},a_{1})$.\qed
\enddemo
Let $0<a<a_{1}$. Let
 $\Sigma_{a}$ be the strip of width $2a$ enclosing $\Gamma$:
$$\Sigma_{a}=\Phi([0,L)\times (-a,a)).$$
Then ${\Bbb R}^{2}\backslash\Sigma_{a}$
 consists of two connected components which
 we denote by $\Lambda^{\text{in}}_{a}$ and
 $\Lambda^{\text{out}}_{a}$,
 where $\Lambda^{\text{in}}_{a}$ is compact.
 We define
$$q_{a,\beta}^{+}(f,f)
=\Vert\nabla f\Vert^{2}_{L^{2}(\Sigma_{a})}
-\beta\int_{\Gamma}|f(x)|^{2}
\,dS\quad
\text{for}\quad
f\in H^{1}_{0}(\Sigma_{a}),$$
$$q_{a,\beta}^{-}(f,f)
=\Vert\nabla f\Vert^{2}_{L^{2}(\Sigma_{a})}
-\beta\int_{\Gamma}|f(x)|^{2}
\,dS\quad\text{for}\quad f\in H^{1}(\Sigma_{a}).$$
Let $L_{a,\beta}^{+}$ and $L_{a,\beta}^{-}$
 be the self-adjoint operators associated with the forms
 $q_{a,\beta}^{+}$ and $q_{a,\beta}^{-}$, respectively.
 By using the Dirichlet-Neumann bracketing (see [7,
 XIII.15, Proposition 4]), we obtain
$$(-\Delta^{\text{N}}_{\Lambda^{\text{in}}_{a}})
\oplus
L_{a,\beta}^{-}
\oplus (-\Delta^{\text{N}}_{\Lambda^{\text{out}}_{a}})
\leq H_{\beta}\leq
(-\Delta^{\text{D}}_{\Lambda^{\text{in}}_{a}})
\oplus
L_{a,\beta}^{+}
\oplus
(-\Delta^{\text{D}}_{\Lambda^{\text{out}}_{a}})
\tag 2.4$$
in $\quad L^{2}(\Lambda^{\text{in}}_{a})
\oplus
L^{2}(\Sigma_{a})\oplus
L^{2}(\Lambda^{\text{out}}_{a})$.
 In order to estimate the negative
 eigenvalues of $H_{\beta}$,
 it is sufficient to estimate those of
 $L_{a,\beta}^{+}$ and $L_{a,\beta}^{-}$ because
 the other operators involved in $(2.4)$ are positive.

To this aim we introduce two operators in
 $L^{2}((0,L)\times (-a,a))$ which are unitarily
 equivalent to $L^{+}_{a,\beta}$ and
 $L^{-}_{a,\beta}$, respectively. We define
$$\align
Q_{a}^{+}=\{\varphi\in H^{1}((0,L)\times (-a,a));\quad
&\varphi(L,\cdot)=\varphi(0,\cdot)
\quad\text{on}\quad (-a,a),\\
&\varphi(\cdot ,a)=\varphi(\cdot, -a)=0
\quad\text{on}\quad (0,L)\},
\endalign
$$
$$Q_{a}^{-}=\{\varphi\in H^{1}((0,L)\times (-a,a));
\quad \varphi(L,\cdot)=
\varphi(0,\cdot)\quad\text{on}\quad
(-a,a)\},$$
$$\align
b^{+}_{a,\beta}(f,f)=&
\int^{L}_{0}\int^{a}_{-a}(1+u\gamma(s))^{-2}
\left|\frac{\partial f}{\partial s}\right|^{2}\,duds
+\int^{L}_{0}\int^{a}_{-a}
\left|\frac{\partial f}{\partial u}\right|^{2}\,duds\\
&+\int^{L}_{0}\int^{a}_{-a}V(s,u)|f|^{2}\,dsdu
-\beta\int^{L}_{0}|f(s,0)|^{2}\,ds
\quad\text{for}\quad
f\in Q^{+}_{a},
\endalign
$$
$$\align
b^{-}_{a,\beta}(f,f)=&
\int^{L}_{0}\int^{a}_{-a}(1+u\gamma(s))^{-2}
\left|\frac{\partial f}{\partial s}\right|^{2}\,duds
+\int^{L}_{0}\int^{a}_{-a}
\left|\frac{\partial f}{\partial u}\right|^{2}\,duds\\
&+\int^{L}_{0}\int^{a}_{-a}V(s,u)|f|^{2}\,dsdu
-\beta\int^{L}_{0}|f(s,0)|^{2}\,ds\\
&-\frac{1}{2}\int^{L}_{0}
\frac{\gamma(s)}{1+a\gamma(s)}|f(s,a)|^{2}\,ds
+\frac{1}{2}\int^{L}_{0}
\frac{\gamma(s)}{1-a\gamma(s)}|f(s,-a)|^{2}\,ds
\endalign
$$
for $f\in Q^{-}_{a}$, where
$$V(s,u)=
\frac{1}{2}(1+u\gamma(s))^{-3}u\gamma^{\prime\prime}(s)
-\frac{5}{4}(1+u\gamma(s))^{-4}u^{2}\gamma^{\prime}(s)^{2}
-\frac{1}{4}(1+u\gamma(s))^{-2}\gamma(s)^{2}.$$
Let $B^{+}_{a,\beta}$ and $B^{-}_{a,\beta}$
 be the self-adjoint operators associated with the forms
 $b^{+}_{a,\beta}$ and $b^{-}_{a,\beta}$, respectively.
 Then we have the following result.
\proclaim{\sc Lemma 2.2} The operators $B^{+}_{a,\beta}$
 and $B^{-}_{a,\beta}$ are unitarily equivalent to
 $L^{+}_{a,\beta}$ and $L^{-}_{a,\beta}$, respectively.
\endproclaim
\demo{Proof} We prove the assertion
 only for $B^{-}_{a,\beta}$
 because that for $B^{+}_{a,\beta}$ is similar.
 Given $f\in L^{2}(\Sigma_{a})$, we define
$$(U_{a}f)(s,u)=(1+u\gamma(s))^{1/2}f(\Phi_{a}(s,u)),\quad
(s,u)\in (0,L)\times (-a,a).\tag 2.5$$
From Lemma 2.1, we
 infer that $U_{a}$ is a unitary operator from
 $L^{2}(\Sigma_{a})$ to $L^{2}((0,L)\times (-a,a))$. Since
 $\Gamma$ is a closed $C^{4}$ Jordan curve, $U_{a}$ is
 a bijection from $H^{1}(\Sigma_{a})$ to $Q^{-}_{a}$.
 Using an integration by parts, we obtain
$$
\align
{}&q^{-}_{a,\beta}(f,g)
-b^{-}_{a,\beta}(U_{a}f,U_{a}g)\\
=&-\frac{1}{2}\int^{a}_{-a}
\left[(1+u\gamma(s))^{-3}\gamma
^{\prime}(s)
(U_{a}f)(s,u)
\overline{(U_{a}g)(s,u)}\right]^{s=a}_{s=-a}
\,du.
\endalign
$$
Since $U_{a}f$ and $U_{a}g$
 as elements of $Q^{-}_{a}$ satisfy the periodicity
 condition, we get
$$q^{-}_{a,\beta}(f,g)
=b^{-}_{a,\beta}(U_{a}f,U_{a}g)\quad\text{for}\quad
f,\,g\in H^{1}(\Sigma_{a}).$$
This together with the first representation theorem
 (see [6, Theorem VI.2.1]) implies that
$$U^{*}_{a}B^{-}_{a,\beta}U_{a}=L^{-}_{a,\beta}.$$
This completes the proof of the lemma.\qed
\enddemo
Next we estimate $B^{+}_{a,\beta}$
 and $B^{-}_{a,\beta}$ by operators with
 separated variables. We put
$$\gamma^{\prime}_{+}
=\max_{[0,L]}|\gamma^{\prime}(\cdot)|,\quad
\gamma^{\prime\prime}_{+}=\max_{[0,L]}|
\gamma^{\prime\prime}(\cdot)|,$$
$$V_{+}(s)
=\frac{1}{2}(1-a\gamma_{+})^{-3}
a\gamma^{\prime\prime}_{+}
-\frac{5}{4}(1+a\gamma_{+})^{-4}
a^{2}(\gamma^{\prime}_{+})^{2}
-\frac{1}{4}(1+a\gamma_{+})^{-2}\gamma(s)^{2},$$

$$V_{-}(s)
=-\frac{1}{2}(1-a\gamma_{+})^{-3}
a\gamma^{\prime\prime}_{+}
-\frac{5}{4}(1-a\gamma_{+})^{-4}
a^{2}(\gamma^{\prime}_{+})^{2}
-\frac{1}{4}(1-a\gamma_{+})^{-2}\gamma(s)^{2}.$$
If $0<a<\frac{1}{2}\gamma_{+}$, we can define
$$\align\tilde{b}^{+}_{a,\beta}(f,f)=&
(1-a\gamma_{+})^{-2}\int^{L}_{0}\int^{a}_{-a}
\left|\frac{\partial f}{\partial s}\right|^{2}\,duds
+\int^{L}_{0}\int^{a}_{-a}
\left|\frac{\partial f}{\partial u}\right|^{2}\,duds\\
&+\int^{L}_{0}\int^{a}_{-a}V_{+}(s)|f|^{2}\,duds
-\beta\int^{L}_{0}|f(s,0)|^{2}\,ds
\quad\text{for}\quad f\in Q^{+}_{a},
\endalign$$
$$\align\tilde{b}^{-}_{a,\beta}(f,f)=&
(1+a\gamma_{+})^{-2}\int^{L}_{0}\int^{a}_{-a}
\left|\frac{\partial f}{\partial s}\right|^{2}\,duds
+\int^{L}_{0}\int^{a}_{-a}
\left|\frac{\partial f}{\partial u}\right|^{2}\,duds\\
&+\int^{L}_{0}\int^{a}_{-a}V_{-}(s)|f|^{2}\,duds
-\beta\int^{L}_{0}|f(s,0)|^{2}\,ds\\
&-\gamma_{+}\int^{L}_{0}(|f(s,a)|^{2}+|f(s,-a)|^{2})\,ds
\quad\text{for}\quad f\in Q^{-}_{a}.
\endalign$$
Then we have
$$b^{+}_{a,\beta}(f,f)\leq \tilde{b}^{+}_{a,\beta}(f,f)
\quad\text{for}\quad f\in Q^{+}_{a},\tag 2.6$$
$$\tilde{b}^{-}_{a,\beta}(f,f)\leq {b}^{-}_{a,\beta}(f,f)
\quad\text{for}\quad f\in Q^{-}_{a}.\tag 2.7$$
Let $\tilde{H}^{+}_{a,\beta}$ and $\tilde{H}^{-}_{a,\beta}$
 be the self-adjoint operators associated
 with the forms $\tilde{b}^{+}_{a,\beta}$
 and $\tilde{b}^{-}_{a,\beta}$
 respectively. Let $T^{+}_{a,\beta}$
 be the self-adjoint operator associated with the form
$$t^{+}_{a,\beta}(f,f)=
\int^{a}_{-a}|f^{\prime}(u)|^{2}\,du-\beta |f(0)|^{2},\quad
f\in H^{1}_{0}((-a,a)).$$
Let finally $T^{-}_{a,\beta}$
 be the self-adjoint operator associated with the form
$$t^{-}_{a,\beta}(f,f)
=\int^{a}_{-a}|f^{\prime}(u)|^{2}\,du-\beta |f(0)|^{2}
-\gamma_{+}(|f(a)|^{2}+|f(-a)|^{2}),
\quad f\in H^{1}((-a,a)).$$
We define
$$U^{+}_{a}=-(1-a\gamma_{+})^{-2}
\frac{d^{2}}{d s^{2}}+V_{+}(s)\quad\text{in}\quad L^{2}((0,L))
\quad\text{with the domain}\quad P,$$
$$U^{-}_{a}=-(1+a\gamma_{+})^{-2}
\frac{d^{2}}{d s^{2}}+V_{-}(s)
\quad\text{in}\quad L^{2}((0,L))
\quad\text{with the domain}\quad P.$$
Then we have
$$\Tilde{H}^{+}_{a,\beta}
=U^{+}_{a}\otimes 1+1\otimes T^{+}_{a,\beta},$$
$$\Tilde{H}^{-}_{a,\beta}
=U^{-}_{a}\otimes 1+1\otimes T^{-}_{a,\beta}.\tag 2.8$$
Next we consider the asymptotic behaviour of each eigenvalue
 of $U^{\pm}_{a}$ as $a$ tends to zero.
 Let $\mu^{\pm}_{j}(a)$ be the
 $j$th eigenvalue of $U^{\pm}_{a}$
 counted with multiplicity.
 The following proposition is needed to prove Theorem 2
 as well as Theorem 1.
\proclaim{\sc Proposition 2.3} There exists $C_{1}>0$ such that
$$|\mu^{+}_{j}(a)-\mu_{j}|\leq C_{1}aj^{2}\tag 2.9$$
and
$$|\mu^{-}_{j}(a)-\mu_{j}|\leq C_{1}aj^{2}\tag 2.10$$
for $j\in{\Bbb N}$ and $0<a<\frac{1}{2\gamma_{+}}$,
 where $C_{1}$ is independent of
 $j$, $a$.
\endproclaim
\demo{Proof} We define
$$S_{0}=-\frac{d^{2}}{ds^{2}}\quad\text{in}
\quad L^{2}((0,L))\quad\text{with the domain}\quad P.$$
Notice that the $j$th eigenvalue of $S_{0}$ counted
 with multiplicity is $4[\frac{j}{2}]^{2}(\frac{\pi}{L})^{2}$.
 Since
$$\Vert S-S_{0}\Vert_{{\Cal B}(L^{2}((0,L)))}
\leq\frac{1}{4}\gamma_{+}^{2},$$
the min-max principle (see [7, Theorem XIII.2]) implies that
$$\left|\mu_{j}-4[j/2]^{2}
(\pi/2)^{2}\right|\leq
\frac{1}{4}\gamma_{+}^{2}\quad\text{for}\quad j\in
{\Bbb N}.\tag 2.11$$
Since
$$
\align
U^{+}_{a}-(1-a\gamma_{+})^{-2}S
=&\frac{1}{2}(1-a\gamma_{+})^{-3}a
\gamma^{\prime\prime}_{+}-\frac{5}{4}
(1+a\gamma_{+})^{-4}a^{2}(\gamma^{\prime}_{+})^{2}\\
&+a\gamma_{+}(1+a\gamma_{+})^{-2}
(1-a\gamma_{+})^{-2}\gamma(s)^{2},
\endalign$$
we infer that there exists $C_{0}>0$ such that
$$\Vert U^{+}_{a}-(1-a\gamma_{+})^{-2}S
\Vert_{{\Cal B}(L^{2}((0,L)))}\leq C_{0}a\quad\text{for}\quad
0<a<\frac{1}{2\gamma_{+}}.$$
This together with the min-max principle implies that
$$|\mu^{+}_{j}(a)-(1-a\gamma_{+})^{-2}\mu_{j}|
\leq C_{0}a\quad\text{for}\quad
0<a<\frac{1}{2\gamma_{+}}.$$
Hence we get
$$|\mu^{+}_{j}(a)-\mu_{j}|\leq
C_{0}a+\frac{a\gamma_{+}(2-a\gamma_{+})}
{(1-a\gamma_{+})^{2}}|\mu_{j}|.$$
Combining this with $(2.11)$ we arrive at $(2.9)$.

 The proof of $(2.10)$ is similar.\qed
\enddemo
Next we estimate the first eigenvalue of $T^{+}_{a,\beta}$.
\proclaim{\sc Proposition 2.4} Assume that
 $\beta a>\frac{8}{3}$.
 Then $T^{+}_{a,\beta}$ has only one negative
 eigenvalue, which we denote by $\zeta^{+}_{a,\beta}$.
 It satisfies the inequalities
$$-\frac{1}{4}\beta^{2}<\zeta^{+}_{a,\beta}<
-\frac{1}{4}\beta^{2}+2\beta^{2}\exp
\left(-\frac{1}{2}\beta a\right).$$
\endproclaim
\demo{Proof} Let $k>0$. We will show that $-k^{2}$
 is an eigenvalue of $T^{+}_{a,\beta}$ if and only if
$$g_{a,\beta}(k):=
\log (\beta-2k)-\log (\beta+2k)+2ka=0.$$
Assume that $-k^{2}$ is an
 eigenvalue of $T^{+}_{a,\beta}$. Notice that
$$\align
{\Cal D}(T^{+}_{a,\beta})=\{\varphi\in H^{1}_{0}((-a,a));
\quad&\varphi|_{(0,a)}\in H^{2}((0,a)),\\
&\varphi|_{(-a,0)}\in H^{2}((-a,0)),\\
&\varphi^{\prime}(+0)-\varphi^{\prime}(-0)
=-\beta\varphi(0)\}.
\endalign
$$
Let a nonzero $\psi$ be
 the eigenfunction of $T^{+}_{a,\beta}$ associated with
 the eigenvalue $-k^{2}$, then we have
\roster
\item"({\bf i})" $-\psi^{\prime\prime}(u)=-k^{2}\psi(u)$
 on $(-a,0)\cup (0,a)$.
\item"({\bf ii})" $\psi(\pm a)=0$.
\item"({\bf iii})" $\psi^{\prime}(+0)-\psi^{\prime}(-0)
=-\beta\psi(0)$.
\endroster

Since $T^{+}_{a,\beta}$ commutes with the parity operator
 $f(x)\mapsto f(-x)$, the ground state $\psi$ satisfies
 $\psi(u)=\psi(-u)$ on $[0,a]$.
 Combining this with ({\bf i}),
 we infer that $\psi$ is of the form
$$\psi(u)=\cases C_{1}e^{ku}+C_{2}e^{-ku},&u\in (0,a),\\
C_{2}e^{ku}+C_{1}e^{-ku},&u\in (-a,0).
\endcases\tag 2.12$$
Note that ({\bf ii})
 is equivalent to
$$C_{2}=-C_{1}e^{2ka}.$$
In addition, ({\bf iii}) is equivalent to
$$(2k+\beta)C_{1}-(2k-\beta)C_{2}=0.$$
Thus the equation for $C_{1}$ and $C_{2}$ becomes
$$\pmatrix
2k+\beta&-(2k-\beta)\\
e^{2ka}&1\endpmatrix
\pmatrix C_{1}\\C_{2}\endpmatrix=0.\tag 2.13$$
Since $(C_{1}, C_{2})\neq (0,0)$, we get
$$\det\pmatrix
2k+\beta&-(2k-\beta)\\
e^{2ka}&1\endpmatrix
=0$$
which is equivalent to $g_{a,\beta}(k)=0$.

To check the converse, assume that $g_{a,\beta}(k)=0$.
 Then $(2.13)$ has a solution $(C_{1}, C_{2})\neq (0,0)$.
 It is easy to see that the function $\psi$ from $(2.12)$
 satisfies ({\bf i})-({\bf iii}) and $\psi\in {\Cal D}
(T^{+}_{a,\beta})$.

Let us show that $g_{a,\beta}(\cdot)$ has
 a unique zero in $(0,\beta/4)$.
 We have $g_{a,\beta}(0)=0$. Since
$$
\frac{d}{dk}g_{a,\beta}(k)=
\frac{-4\beta}{\beta^{2}-4k^{2}}+2a,$$
we claim that $g_{a,\beta}(\cdot)$ is monotone increasing
 on $(0,\frac{1}{2}\sqrt{\beta^{2}-\frac{2\beta}{a}})$
 and is monotone decreasing on $(\frac{1}{2}
\sqrt{\beta^{2}-\frac{2\beta}{a}},\frac{1}{2}\beta)$.
 Moreover, we have
$$\lim_{k\to\frac{\beta}{2}-0}g_{a,\beta}(k)=-\infty.$$
Hence the function $g_{a,\beta}(\cdot)$
 has a unique zero in $(0, \,\beta/2)$.
 Since $a\beta>\frac{8}{3}$, we have
 $\frac{1}{2}\sqrt{\beta^{2}-\frac{2\beta}{a}}
\geq\frac{\beta}{4}$.
 Consequently, the solution $k$
 has the form $k=\frac{\beta}{2}-s$,
 $0<s\leq\frac{\beta}{4}$.
 Taking into account the relation $g_{a,\beta}(k)=0$, we get
$$\log 2s=\log (2\beta-s)-\beta a+2as
<\log 2\beta-\frac{1}{2}a\beta.$$
So we obtain $s<\beta\exp(-\frac{1}{2}a\beta)$.
 This completes the proof of Proposition 2.4.\qed
\enddemo
Next we estimate the first eigenvalue of $T^{-}_{a,\beta}$.
\proclaim{\sc Proposition 2.5} Let $a\beta>8$
 and $\beta>\frac{8}{3}\gamma_{+}$.
 Then $T^{-}_{a,\beta}$ has
 a unique negative eigenvalue $\zeta^{-}_{a,\beta}$,
 and moreover, we have
$$-\frac{1}{4}\beta^{2}-
\frac{2205}{16}\beta^{2}\exp\left(-\frac{1}{2}\beta a\right)
<\zeta^{-}_{a,\beta}<-\frac{1}{4}\beta^{2}.$$
\endproclaim
\demo{Proof} Let us first show that $T^{-}_{a,\beta}$
 has a unique negative eigenvalue. Let $k>0$.
 As in the proof
 of Proposition 2.4, we infer that $-k^{2}$
 is an eigenvalue of $T^{-}_{a,\beta}$ if and only if
$$\frac{ke^{ka}-\gamma_{+}}{ke^{-ka}+\gamma_{+}}
=\frac{2k+\beta}{2k-\beta}.\tag 2.14$$
Since the left side of $(2.14)$
 is positive for $k\geq\gamma_{+}$
 and the right side of
 $(2.14)$ is negative for $0<k<\frac{\beta}{2}$,
 $(2.14)$ has no solution in $[\gamma_{+},\frac{\beta}{2})$.
 We put
$$g(k)=\frac{ke^{ka}-\gamma_{+}}{ke^{-ka}+\gamma_{+}}
\quad\text{and}\quad h(k)=\frac{2k+\beta}{2k-\beta}.$$
Then we get $\lim_{k\rightarrow\infty}g(k)=\infty$ and
$$g^{\prime}(k)=
\frac{\gamma_{+}(e^{ka}-e^{-ka})+2k^{2}a
+ka\gamma_{+}(e^{ka}+e^{-ka})}
{(ke^{-ka}+\gamma_{+})^{2}}>0\quad\text{for}\quad k>0.$$
Thus $g(k)$ is monotone increasing on $(0,\infty)$.
 On the other hand, $h(k)$
 is monotone decreasing on $(\beta/2,\infty)$,
$$\lim_{k\to\frac{\beta}{2}+0}h(k)=\infty,\quad
\lim_{k\to\infty}h(k)=1.$$
Hence $(2.14)$ has a unique solution in $(\beta/2,\infty)$.
 Since $h(k)$ is monotone decreasing on $(0,\beta/2)$
 and $g(0)=h(0)$, we claim that $(2.14)$
 has no solution in $(0,\beta/2)$.

Next we show that $g(k)>\frac{2k+\beta}{2k-\beta}$
 for $k\geq\frac{3}{4}\beta$. We have
 $\frac{2k+\beta}{2k-\beta}\leq 5$ for
 $k\geq\frac{3}{4}\beta$.
 For $k\geq\frac{3}{4}\beta$, we get
$$\align
g(k)&\geq g\left(\frac{3}{4}\beta\right)\\
&=\frac{
\frac{3}{4}\beta\exp({\tsize\frac{3}{4}}a\beta)-\gamma_{+}
}
{
\frac{3}{4}\beta\exp(-\frac{3}{4}a\beta)+\gamma_{+}
}\\
\intertext{since $\gamma_{+}<\frac{3}{8}\beta
<\frac{3}{8}\beta\exp(\frac{3}{4}a\beta)$}
&\geq\frac{
\frac{3}{8}\beta\exp(\frac{3}{4}a\beta)
}
{
\frac{3}{4}\beta\exp(-\frac{3}{4}a\beta)
+\frac{3}{8}\beta
}\\
&=\frac{\exp(\frac{3}{4}a\beta)}
{2\exp(-\frac{3}{4}a\beta)+1}\\
\intertext{since $a\beta>8$}
&\geq \frac{e^{6}}{2e^{-6}+1}\\
&>5.
\endalign
$$
So $(2.14)$ has no solution in $[\frac{3}{4}\beta,\infty)$.
 Hence, the solution $k$ of $(2.14)$ is of the form
 $k=\frac{\beta}{2}+s$, $0<s<\frac{1}{4}\beta$.
 From $(2.14)$, we get
$$\align
\frac{5\beta}{4s}&\geq\frac{2k+\beta}{2k-\beta}\\
&=\frac{ke^{ka}-\gamma_{+}}{ke^{-ka}+\gamma_{+}}\\
\intertext{since $\gamma_{+}<\frac{3}{8}\beta
<\frac{3}{8}\beta\exp(\frac{1}{2}\beta a)$ and
 $ke^{ka}\geq\frac{1}{2}\beta\exp(\frac{1}{2}\beta a)$}
&\geq\frac{\frac{1}{8}\beta\exp(\frac{1}{2}\beta a)
}{ke^{-ka}+\gamma_{+}}\\
\intertext{since $ke^{-ka}<k<\frac{3}{4}\beta$
 and $\gamma_{+}<\frac{3}{8}\beta$}
&\geq\frac{\frac{1}{8}\beta\exp(\frac{1}{2}\beta a)}
{\frac{9}{8}\beta}\\
&=
{\tsize\frac{1}{9}}\exp({\tsize\frac{1}{2}}\beta a).
\endalign
$$
Thus we get $s\leq\frac{45}{4}\beta\exp(-\frac{1}{2}\beta a)$,
 which gives $k^{2}\geq\frac{\beta^{2}}{4}$ and
$$\align
k^{2}=&\frac{\beta^{2}}{4}+\beta s+s^{2}\\
\leq&\frac{\beta^{2}}{4}+\frac{45}{4}\beta^{2}
\exp\left(-\frac{1}{2}\beta a\right)
+\left(\frac{45}{4}\right)^{2}\beta^{2}\exp\left(-\beta a\right)\\
\leq&\frac{\beta^{2}}{4}+\frac{45}{4}\beta^{2}
\exp\left(-\frac{1}{2}\beta a\right)
+\left(\frac{45}{4}\right)^{2}\beta^{2}\exp\left(
-\frac{1}{2}\beta a\right)\\
=&\frac{\beta^{2}}{4}+\frac{2205}{16}\exp\left(
-\frac{1}{2}\beta a\right).
\endalign$$
This completes the proof of Proposition 2.5.\qed
\enddemo

Now we are ready to prove Theorem 1.
\demo{Proof of Theorem 1} We put $a(\beta)
=6\beta^{-1}\log\beta$. Let $\xi^{\pm}_{\beta,j}$
 be the $j$th eigenvalue of $T^{\pm}_{a(\beta),\beta}$.
 From Propositions 2.4 and 2.5, we have
$$\xi^{\pm}_{\beta,1}
=\zeta^{\pm}_{a(\beta),\beta}\quad\text{and}\quad
\xi^{\pm}_{\beta,2}\geq 0.$$
From $(2.8)$, we infer that
 $\{\xi^{\pm}_{\beta,j}
+\mu^{\pm}_{k}(a(\beta))\}_{j,k\in{\Bbb N}}$ is
 a sequence of all eigenvalues of
 $\tilde{H}^{\pm}_{a(\beta),\beta}$
 counted with multiplicity. From Proposition
 2.3, we have
$$\xi^{\pm}_{\beta,j}
+\mu^{\pm}_{k}(a(\beta))\geq\mu^{\pm}_{1}(a(\beta))
=\mu_{1}+{\Cal O}(\beta^{-1}\log\beta)\tag 2.15$$
for $j\geq 2$ and $k\geq 1$.
 For $j\in{\Bbb N}$, we define
$$\tau^{\pm}_{\beta,j}
=\zeta^{\pm}_{a(\beta),\beta}
+\mu^{\pm}_{j}(a(\beta)).\tag 2.16$$
From Propositions 2.3--2.5, we get
$$\tau^{\pm}_{\beta,j}=
-\frac{1}{4}\beta^{2}
+\mu_{j}+{\Cal O}(\beta^{-1}\log\beta)
\quad\text{as}\quad\beta\rightarrow\infty.\tag 2.17$$
Let $n\in {\Bbb N}$. Combining $(2.15)$ with $(2.17)$,
 we claim that there exists $\beta(n)>0$ such that
$$
\tau^{+}_{\beta,n}<0,
\quad\tau^{+}_{\beta,n}<\xi^{+}_{\beta,j}
+\mu^{+}_{k}(a(\beta)),\quad\text{and}\quad
\tau^{-}_{\beta,n}<\xi^{-}_{\beta,j}
+\mu^{-}_{k}(a(\beta))$$
for $\beta\geq\beta(n)$, $j\geq 2$, and $k\geq 1$.
 Hence the $j$th eigenvalue of
 $\tilde{H}^{\pm}_{a(\beta),\beta}$
 counted with multiplicity is $\tau^{\pm}_{\beta,j}$
 for $j\leq n$ and $\beta\geq\beta(n)$.
 Let $\beta\geq \beta(n)$ and denote by
 $\kappa^{\pm}_{j}(\beta)$ the $j$th eigenvalue of
 $L^{\pm}_{a(\beta),\beta}$. From $(2.4)$, $(2.6)$, and
 the min--max principle we obtain
$$\tau^{-}_{\beta,j}\leq\kappa^{-}_{j}(\beta)
\quad\text{and}\quad\kappa^{+}_{j}(\beta)\leq
\tau^{+}_{\beta,j}\quad\text{for}\quad 1\leq j\leq n,
\tag 2.18$$
so we have $\kappa^{+}_{n}(\beta)<0$.
 Hence the min--max principle and $(2.4)$ imply that
 $H_{\beta}$ has at least $n$ eigenvalues
 in $(-\infty,\kappa^{+}_{n}(\beta))$.
 For $1\leq j\leq n$, we denote by
 $\lambda_{j}(\beta)$ the $j$th eigenvalue
 of $H_{\beta}$. We have
$$\kappa^{-}_{j}(\beta)\leq
\lambda_{j}(\beta)\leq\kappa^{+}_{j}(\beta)
\quad\text{for}\quad
1\leq j\leq n.$$
This together with $(2.17)$ and $(2.18)$ implies that
$$\lambda_{j}(\beta)=-\frac{1}{4}\beta^{2}
+\mu_{j}+{\Cal O}(\beta^{-1}\log\beta)
\quad\text{as}\quad
\beta\rightarrow\infty\quad\text{for}\quad 1\leq j\leq n.$$
This completes the proof of Theorem 1.\qed
\enddemo
\leftline{}
\leftline{\bf 3. Proof of Theorem 2}
\leftline{}
\noindent
For a self-adjoint operator $A$, we define
$$N^{-}(A)=\sharp
 \{\sigma_{\roman d}(A)\cap (-\infty,0)\}.$$
From $(2.4)$, we have
 $N^{-}(L^{-}_{a,\beta})\geq \sharp
\sigma_{\roman d}(H_{\beta})
\geq N^{-}(L^{+}_{a,\beta})$.
 On the other hand, Lemma 2.2,
 $(2.6)$, and $(2.7)$ imply that
 $N^{-}(\tilde{H}^{-}_{a,\beta})
\geq N^{-}(L^{-}_{a,\beta})$ and
 $N^{-}(L^{+}_{a,\beta})
\geq N^{-}(\tilde{H}^{+}_{a,\beta})$. In this way we get
$$N^{-}(\tilde{H}^{+}_{a,\beta})\leq
 \sharp\sigma_{\roman d}(H_{\beta})
\leq N^{-}(\tilde{H}^{-}_{a,\beta}).\tag 3.1$$
Recall the relation $(2.16)$. We define
$$K^{\pm}_{\beta}=\{j\in {\Bbb N};
\quad\tau^{\pm}_{\beta,j}<0\}$$
and use the following proposition to estimate $N^{-}(\tilde{H}
^{\pm}_{a,\beta})$.
\proclaim{\sc Proposition 3.1} We have
$$\sharp K^{\pm}_{\beta}
=\frac{L}{2\pi}\beta
+{\Cal O}(\log\beta)
\quad\text{as}\quad\beta\rightarrow\infty.$$
\endproclaim
\demo{Proof} We choose $C_{2}>0$ such that
 $-\frac{1}{4}C_{2}^{2}\leq
-1-\frac{1}{4}\gamma_{+}^{2}$. Let
 $\beta\geq\max\{ 2,\, C_{2}\}$. Then we have
 $\frac{1}{4}(\beta-C_{2})^{2}<\frac{1}{4}\beta^{2}
-1-\frac{1}{4}\gamma_{+}^{2}$. We get
$$\align
K^{+}_{\beta}=&\{j\in {\Bbb N};
\quad\mu^{+}_{j}(a(\beta))
<-\zeta^{+}_{a(\beta),\beta}\}\\
\intertext{by using Propositions 2.3 and 2.4}
\supset&\{j\in {\Bbb N};\quad \mu_{j}
+C_{1}a(\beta)j^{2}<{\tsize\frac{1}{4}}
\beta^{2}-2\beta^{2}
\exp(-{\tsize\frac{1}{2}}\beta a(\beta))\}\\
\intertext{since $\mu_{j}\leq [j/2]^{2}
(\pi/L)^{2}+\frac{1}{4}\gamma_{+}^{2}$}
\supset&\{j\in {\Bbb N};\quad 4[j/2]^{2}
(\pi/L)^{2}+C_{1}(\beta^{-1}\log\beta)j^{2}<
{\tsize\frac{1}{4}}\beta^{2}
-{\tsize\frac{2}{\beta}}
-{\tsize\frac{1}{4}}\gamma_{+}^{2}\}\\
\intertext{since $\beta\geq 2$}
\supset&\{j\in {\Bbb N};\quad j^{2}(\pi/L
)^{2}+C_{1}(\beta^{-1}\log\beta)j^{2}
<{\tsize\frac{1}{4}}\beta^{2}
-1-{\tsize\frac{1}{4}}\gamma_{+}^{2}\}\\
\supset&\{j\in {\Bbb N};\quad j^{2}(\pi/L)^{2}
+C_{1}(\beta^{-1}\log\beta)j^{2}\leq
{\tsize\frac{1}{4}}(\beta-C_{2})^{2}\}\\
=&\{j\in {\Bbb N};\quad j
\leq {\tsize\frac{1}{2}}(\beta-C_{2})
((\pi/L)^{2}+C_{1}\beta^{-1}\log\beta)^{-1/2}\}.
\endalign
$$
Furthermore, from
$${\tsize\frac{1}{2}}(\beta-C_{2})
((\pi/L)^{2}+C_{1}\beta^{-1}\log\beta)^{-1/2}
={\tsize\frac{L\beta}{2\pi}}
+{\Cal O}(\log\beta)\quad\text{as}
\quad\beta\rightarrow\infty,$$
we infer that
$$\sharp K^{+}_{\beta}\geq
{\tsize\frac{L\beta}{2\pi}}
+{\Cal O}(\log\beta)
\quad\text{as}\quad
\beta\rightarrow\infty.\tag 3.2$$
Similarly we get
$$\align
K^{-}_{\beta}=&\{j\in {\Bbb N};\quad \mu^{-}_{j}(a(\beta))
<-\zeta^{-}_{a(\beta),\beta}\}\\
\supset&\{j\in {\Bbb N};\quad \mu_{j}-C_{1}
a(\beta)j^{2}<
{\tsize\frac{1}{4}}\beta^{2}
+{\tsize\frac{2205}{4\beta}}+
{\tsize\frac{1}{4}}\gamma_{+}^
{2}\}\\
\intertext{since $2(j-1)\geq j$
 for $j\geq 2$}
=&\{1\}\cup\{j\geq 2;\quad (j-1)^{2}
(\pi/L)^{2}-4C_{1}(\beta^{-1}\log\beta)(j-1)^{2}<
{\tsize\frac{1}{4}}\beta^{2}
+{\tsize\frac{2205}{4\beta}}
+{\tsize\frac{1}{4}}\gamma_{+}^
{2}\}\\
=&\{1\}\cup\{j\geq 2;\quad j
<1+({\tsize\frac{1}{4}}\beta^{2}
+{\tsize\frac{2205}{4\beta}}
+{\tsize\frac{1}{4}}\gamma_{+}^
{2})((\pi/L)^{2}-4C_{1}
\beta^{-1}\log\beta)^{-1/2}\}.
\endalign$$
However,
$$1+({\tsize\frac{1}{4}}\beta^{2}+
{\tsize\frac{2205}{4\beta}}
+{\tsize\frac{1}{4}}\gamma_{+}^
{2})((\pi/L
)^{2}-4C_{1}\beta^{-1}\log\beta)^{-1/2}
={\tsize\frac{L\beta}{2\pi}}
+{\Cal O}(\log\beta)$$
as $\beta\rightarrow\infty$,
 which leads to
$$\sharp K^{-}_{\beta}\leq
{\tsize\frac{L\beta}{2\pi}}
+{\Cal O}(\log\beta)
\quad\text{as}\quad
\beta\rightarrow\infty.\tag
3.3$$
Since $\tau^{-}_{\beta,j}
<\tau^{+}_{\beta,j}$, we get
 $K^{-}_{\beta}\supset K^{+}_{\beta}$.
Combining this with $(3.2)$ and $(3.3)$,
 we get the assertion of Proposition 3.1.\qed
\enddemo
We also need the following result
 to estimate the second eigenvalue
 of $T^{-}_{a,\beta}$.
\proclaim{\sc Proposition 3.2} Let
 $0<a<\frac{1}{\sqrt{2}\gamma_{+}}$
 and $\beta>0$.
 Then $T^{-}_{a,\beta}$ has no eigenvalue
 in $[0,
\min\{\frac{\pi^{2}}{16a^{2}},
\frac{\beta\gamma_{+}}{2},\beta^{2}\})$.
\endproclaim
\demo{Proof} Let $k>0$. As in the proof
 of Proposition 2.4, we infer that
 $k^{2}$ is an eigenvalue of
 $T^{-}_{a,\beta}$ if and only if $k$ solves either
$$\tan ka={\tsize\frac{k}{\gamma_{+}}}\tag 3.4$$
or
$$\tan ka={\tsize\frac{\beta+2k\gamma_{+}}
{\beta\gamma_{+}-2k^{2}}}\beta.\tag 3.5$$
For $k\in (0,\frac{\pi}{4a})$, we have
$$\tan ka<\sqrt{2}\sin ka<\sqrt{2}ka
<{\tsize\frac{k}{\gamma_{+}}}.\tag 3.6$$
Thus $(3.4)$ has no solution in
 $(0,\frac{\pi}{4a})$. For
 $k\in (0,
\min\{\frac{\pi}{4a},
\frac{\sqrt{\beta\gamma_{+}}}{\sqrt{2}},
\beta\})$, we have
$$\frac{\beta+2k\gamma_{+}}{\beta\gamma_{+}
-2k^{2}}\beta-\frac{k}{\gamma_{+}}
=\frac{\beta\gamma_{+}(\beta-\gamma_{+})
+2k(\gamma_{+})^{2}\beta+2k^{3}}
{(\beta\gamma_{+}-2k^{2})\gamma_{+}}>0.$$

This together with $(3.6)$ implies
 that $(3.5)$ has no solution
 in
 $(0,
\min\{\frac{\pi}{4a},
\frac{\sqrt{\beta\gamma_{+}}}{\sqrt{2}},
\beta\})$. Consequently, $T^{-}_{a,\beta}$ has no eigenvalue
 in $(0,
\min\{\frac{\pi^{2}}{16a^{2}},
\frac{\beta\gamma_{+}}{2},\beta^{2}\})$.

Next we show that $0$ is not an eigenvalue of $T^{-}_{a,\beta}$.
 As in the proof of Proposition 2.4, we infer that $0$ is an
 eigenvalue of $T^{-}_{a,\beta}$ if and only if either
 $\gamma_{+}a=1$ or $\beta(\gamma_{+}a-1)=2\gamma_{+}$ holds. Since
 $0<a<\frac{1}{\sqrt{2}\gamma_{+}}$ and $\beta>0$, we have
 $\gamma_{+}a<1$ and $\beta(\gamma_{+}a-1)<2\gamma_{+}$. Hence
 $0$ is not an eigenvalue of $T^{-}_{a,\beta}$,
 and the proof is
 complete.\qed
\enddemo
Now we are in a position to prove Theorem 2.
\demo{Proof of Theorem 2} Let us first show that
$$N^{-}(\tilde{H}^{-}_{a(\beta),\beta})
=\sharp K^{-}_{\beta}\quad\text{for
 sufficiently large}\quad\beta>0.
\tag 3.7$$
Recall
 that $\{\xi^{-}_{\beta,j}+\mu^{-}_{k}(a(\beta))\}
_{j,k\in{\Bbb N}}$ is a sequence of all
 eigenvalues of $\tilde{H}^{-}_{a(\beta),\beta}$
 counted with multiplicity.
 From Proposition 3.2, we have
$$\xi^{-}_{\beta,2}\geq
\min\left\{\frac{\pi}{4a(\beta)},
\frac{\sqrt{\beta\gamma_{+}}}{\sqrt{2}},
\beta\right\}.$$
This together with $(2.10)$ implies that
 there exists $\beta_{0}>0$ such that $\xi^{-}_{\beta,2}
+\mu^{-}_{1}(a(\beta))>0$ for $\beta\geq\beta_{0}$.
 We obtain
$$\xi^{-}_{\beta,j}+\mu^{-}_{k}(a(\beta))>0
\quad\text{for}\quad j\geq 2,\quad k\geq 1,\quad
\text{and}\quad\beta\geq\beta_{0}.$$
Thus we get
$$\align
N^{-}(\tilde{H}^{-}_{a(\beta),\beta})
=&\sharp\{(j,k)\in {\Bbb N}^{2};
\quad \xi^{-}_{\beta,j}+\mu^{-}_{k}(a(\beta))<0\}\\
=&\sharp\{j\in {\Bbb N};\quad\tau^{-}_{\beta,j}<0\}\\
=&\sharp K^{-}_{\beta}
\quad\text{for}\quad\beta\geq\beta_{0}.
\endalign
$$
In this way we obtain $(3.7)$.
From $(3.1)$, we get
$$\sharp K^{+}_{\beta}\leq
\sharp\sigma_{\roman d}(H_{\beta})
\leq N^{-}(\tilde{H}^{-}_{a(\beta),\beta}).$$
This together with $(3.7)$ and Proposition 3.1
 implies the assertion of Theorem 2.\qed
\enddemo
\remark{\sc Remark 3.3} We can also prove $(1.5)$
 in the case that $\gamma$ is a $C^{4}$ curve which is not
 self-intersecting. Indeed, it suffices to use
 the following operators $\hat{H}^{\pm}_{a,\beta}$
 instead of $\tilde{H}_{a,\beta}=U^{\pm}_{a}
\otimes1+1\otimes T^{\pm}_{a,\beta}$:
$$\hat{H}_{a,\beta}:=
\hat{U}^{\pm}_{a}\otimes 1+1\otimes T^{\pm}_{a,\beta}
\quad\text{in}\quad
L^{2}((0,L))\otimes L^{2}((-a,a))
=L^{2}((0,L)\times(-a,a)),$$
$$\hat{U}^{+}_{a}:=
-(1-a\gamma_{+})^{-2}\frac{d^{2}}{ds^{2}}+V_{+}(s)
\quad\text{in}\quad L^{2}((0,L))$$
with the
 Dirichlet boundary condition,
$$\hat{U}^{-}_{a}:=
-(1+a\gamma_{+})^{-2}\frac{d^{2}}{ds^{2}}+V_{-}(s)
\quad\text{in}\quad L^{2}((0,L))$$
with the
 Neumann boundary condition.
\endremark
\leftline{}
\leftline{\ttt Acknowledments.}
\leftline{}
\noindent The research has been partially
supported by GAAS and the Czech Ministry of Education under the
projects A1048101 and ME170.
\leftline{}
\leftline{}

\enddocument